\documentclass[10pt,letterpaper,twocolumn]{article} 

\usepackage{ol2}
\usepackage{bm,braket}
\usepackage{float}
\usepackage{graphicx}
\usepackage{amsfonts}
\usepackage{amsmath}
\usepackage{amssymb}
\usepackage{color}
\usepackage{ulem}
\newcommand {\be}{\begin{equation}}
\newcommand {\ee}{\end{equation}}
\newcommand {\bea}{\begin{eqnarray}}
\newcommand {\eea}{\end{eqnarray}}
\newcommand {\nn}{\nonumber}
\renewcommand{\v}[1]{\ensuremath{\mathbf{#1}}} 

\newcommand{\avg}[1]{\left< #1 \right>} 

\def\sm{\sigma^-}
\def\smdag{\sigma^+}
\def\f{\textbf{f}}
\def\fdag{\textbf{f}^\dagger}

\def\r1{\textbf{r}}

\begin{document}
\abovedisplayskip=6pt
\abovedisplayshortskip=6pt
\belowdisplayskip=6pt
\belowdisplayshortskip=6pt
\twocolumn[ 
\title{Spontaneous emission from a quantum dot
in a structured photonic reservoir: phonon-mediated breakdown of Fermi's golden rule}
\author{Kaushik Roy-Choudhury and Stephen Hughes}

\affiliation{Department of Physics, Queen's University, Kingston, Ontario, Canada, K7L 3N6}



\begin{abstract}

Quantum dots in semiconductor photonic reservoirs are important systems for studying and exploiting quantum optics on a chip, and it is essential to understand fundamental concepts such as spontaneous emission. According to Fermi's golden rule, the spontaneous emission rate of a quantum emitter weakly coupled to a structured photonic reservoir is proportional to the local density of photon states (LDOS) at the emitter's position and frequency. Coupling to lattice vibrations or phonons however significantly modifies the emission properties of a quantum dot compared to an isolated emitter (e.g., an atom). In the regime of phonon-dressed reservoir coupling, we demonstrate why and how the broadband frequency dependence of the LDOS determines the spontaneous emission rate of a QD, manifesting in a  dramatic breakdown of Fermi's golden rule. We analyze this problem using a polaron transformed master equation and consider specific examples of a semiconductor  microcavity and a coupled cavity waveguide. For a leaky single cavity resonance, we generalize Purcell's formula to include the effects of electron-phonon coupling and for a waveguide, we show  a suppression and  a 200-fold enhancement of the  photon emission rate. These results have important consequences for modelling and understanding emerging QD experiments in a wide range of photonic reservoir systems.

\end{abstract}

\ocis{(270.0270); (350.4238); (160.6000).}

] 





\maketitle



Recent developments in chip-scale quantum  optical technologies~\cite{Kim} have generated substantial interest in quantum dots (QDs)  which act as ``artificial atoms'' in solid-state media.
However, electron-phonon coupling in solid-state media has been shown to significantly modify the emission properties of a QD as compared to an isolated atom~\cite{Weiler}. Studying  phonon interactions in governing the emission properties of QDs has been an intense area of research, leading to a number of  effects beyond a simple pure-dephasing model~\cite{Axt2}. For driven QD excitons yielding Rabi oscillations, phonon coupling manifests in  damping and frequency shifts  ~\cite{Forstner, Ramsay, Leonard}. In  QD-cavity systems, phonons  cause intensity-dependent  broadening of Mollow side-bands~\cite{Ulrich}, off-resonant cavity feeding~\cite{Arka} and asymmetric vacuum Rabi doublets~\cite{Milde,Ota}.

\begin{figure}[t]
\includegraphics[width=0.99\columnwidth]{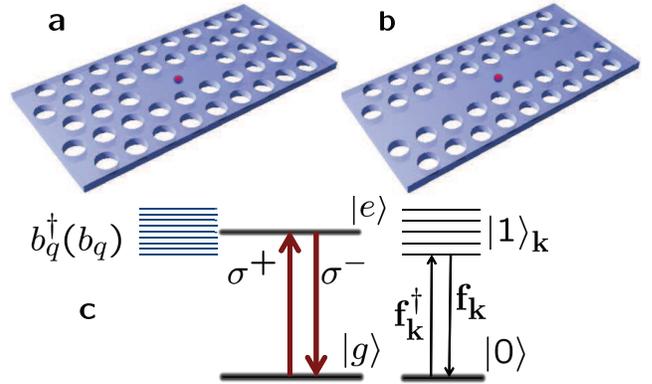}
\vspace{-0.cm}
\caption{\label{fig1} \footnotesize{  {\bf  Two photonic reservoir systems under consideration and an energy level picture of the various quantum states.}
Schematic of a semiconductor cavity ({\bf a}) and
waveguide ({\bf b}) using a photonic crystal platform, containing a single QD. ({\bf c}) Energy level diagram of a neutral QD  (electron-hole pair) interacting with a phonon bath and a photon bath. The operator ${\bf f}^\dag_{\bf k}$ creates a photon}
and the $b^{\dag}_q$ operator creates a phonon.}
\label{fig:1}
\end{figure}

Semiconductor quantum dots (QDs) coupled to structured photonic reservoirs, provide a promising platform for tailoring light-matter interaction in a solid-state environment. One of the primary interests in coupling QDs to structured reservoirs is for modifying the spontaneous emission rate (SE), $\gamma$, via the Purcell effect~\cite{Purcell,Dirk}. Photonic crystals are a paradigm example of a structured photonic reservoir, and both photonic crystal cavities (Fig.\ref{fig:1}(a)) and coupled-cavity  optical waveguide (CROW, Fig.\ref{fig:1}(b)) 
structures have been investigated for modifying QD~SE rates \cite{Dirk, Lodahl}. For an unstructured reservoir, $\gamma$ remains unchanged in the presence of phonons~\cite{Nazir}. 
For structured reservoirs, previous theories have  assumed phonon processes to be much faster than all relevant system dynamics~\cite{Kaer, Roy1}, thus restricting them to structures with sharp variations of photon local density of states (LDOS) (e.g., high-$Q$ cavity or photonic band edge). A primary example of a structured reservoir is a microcavity, and existing theories~\cite{Roy2, Kaer, Hohenester} treat the cavity mode as a system
operator and find that phonons modify the QD-cavity coupling rate, through $g\rightarrow \braket{B}\!g$ \cite{Roy2, Imamoglu}, where $\avg{B}$ is the thermal average of the coherent phonon bath displacement operators $B_{\pm}$~\cite{Roy2}.
Hence the Purcell factor is believed to scale as $g^2\rightarrow \braket{B}^2\!g^2$
~\cite{Roy1}.
However, such theories do not apply to large $\kappa$ cavities,
where $\kappa$ is the cavity decay rate, and one would expect to recover the result that $\gamma$---and thus $g$---are not affected by phonons. Moreover, for an arbitrary photonic bath medium,
it is not known  how phonons affect the SE rates, yet clearly such an effect is of significant fundamental interest and also important for  understanding  emerging QD experiments.

In this Letter we
introduce a self-consistent ME approach with both phonon and photon reservoirs included and we explore in detail the influence of a photon reservoir on the phonon-modified SE rate.
When the relaxation times of the photon and phonon baths compare, the frequency dependence of the LDOS is found to dictate
how phonons modify the SE rates, causing a clear breakdown of Fermi's golden rule. Such an effect arises due to dressing of QD excitations by phonons in a solid state medium.
 Importantly, 
our theory can be applied to any general LDOS medium, and as specific examples we consider   a semiconductor microcavity and a slow-light coupled cavity  waveguide.

We model the QD as a two-level system interacting with an   inhomogeneous semiconductor-based photonic reservoir and an acoustic phonon bath~\cite{Roy2} (see Fig.~\ref{fig:1} (c)). Assuming the QD of dipole moment $\v{d}=d\v{\hat n}_d$  at spatial position $\r1_d$, the total Hamiltonian of the system in a frame rotating at the QD exciton frequency $\omega_x$, is~\cite{Scheel}
\begin{align}
\label{eq1}
& H = \hbar\int d\v{r} \int_0^{\infty}d\omega\,\fdag(\r1,\omega)\f(\r1,\omega) + \Sigma_q\hbar \omega_q b^{\dag}_qb_q  \nn\\
&-\left[\smdag e^{i\omega_xt}\int_0^{\infty}d\omega\,\v{d}\cdot\v{E}(\r1_d,\omega) + \text{H.c}\right] 
\nn\\
&+ \smdag\sm\Sigma_q\hbar\lambda_q(b^{\dag}_q+b_q),
\end{align}
where $\smdag$/$\sm$ are the Pauli operators of the exciton 
(electron-hole pair), $b_q$/$b^{\dag}_q$ are the annihilation and creation operators of the acoustic phonon reservoir and $\lambda_q$ is the exciton-phonon coupling strength. The operators $\f$/$\fdag$ are the boson field operators of the photon reservoir and these satisfy the usual commutation rules for boson operators. The interaction between the QD and the photonic reservoir is written using the dipole and the rotating wave approximations. The electric-field operator $\v{E}(\r1,\omega)$ is 
related to the  Green function of the medium, $\v{G}(\r1,\r1';\omega)$,
which satisfies the Kramers-Kronig relations~\cite{Scheel}.
To include phonon interactions to all orders, we  perform the polaron transform on the Hamiltonian $H$ given by $H' \rightarrow e^P He^{-P}$ where $P = \smdag\sm \Sigma_q\frac{\lambda_q}{\omega_q} (b^{\dag}_q-b_q)$~\cite{Imamoglu}. %
Assuming weak-to-intermediate coupling with the photon bath, we derive a  time-local polaron ME for the QD reduced density operator $\rho$,  using the Born approximation. The usual incoherent terms from the photon
reservoir can be written as $\frac{\partial \rho}{\partial t} |^{ph}_{inc} = {\cal L}_{ph}(\rho)$. Subsequently,  the phonon-modified SE decay rate can be obtained from  $\text{Re}(\mathcal{L}_{\text{ph}})$,
 yielding  the familiar Lindblad superoperator, $\tilde{\gamma}(t) \frac{1}{2}(2\sm\rho\smdag -\smdag\sm \rho-\rho \smdag\sm)$, where the SE decay rate is derived to be (See Supplement 1, Sec. 2)
\begin{align}
\tilde{\gamma}(t) = 2\int_0^t\text{Re}[C_{\text{pn}}(\tau)J_{\text{ph}}(\tau)]d\tau,
\label{eq:SE}
\end{align}

where $J_{\text{ph}}(\tau)$ and $C_{\text{pn}}(\tau)$ are the photon and the phonon bath correlation functions, respectively. The {phonon} bath correlation function $C_{\text{pn}}(\tau)$ is defined as $C_{\text{pn}}(\tau)=e^{[\phi(\tau)-\phi(0)]}$ where 
$\phi(t) = \int_0^{\infty} d\omega\frac{J_{\rm pn}(\omega)}{\omega^2}[\coth(\hbar\omega/2k_BT)\cos(\omega t)- i\sin(\omega t)]$, and $J_{\rm pn}(\omega)$ is the {phonon} spectral function~\cite{Roy3}. The photon bath correlation function can be expressed in terms of the photon-reservoir spectral function $J_{\text{ph}}(\omega)= \frac{\v{d}\cdot \text{Im}[\v{G}(\r1_d,\r1_d;\omega)]\cdot \v{d}}{\pi\hbar\epsilon_0}$, with
$J_{\text{ph}}(\tau) = \int_0^{\infty} d\omega J_{\text{ph}}(\omega) e^{i(\omega_x-\omega)\tau}$. 
In the Markov limit ($t\rightarrow \infty$), equation~(\ref{eq:SE}) generalizes Fermi's golden rule for QD SE, since the LDOS at various frequencies can now contribute to the phonon-modified SE rate,   $\tilde{\gamma}$. Similar expressions for the SE rate in the frequency domain have been used to explain mode pulling effects in QD cavities~\cite{Valente}. In the absence of phonon coupling, the SE decay rate of the QD in a structured photon reservoir reduces to $\gamma(t) = 2\int_0^t\text{Re}[J_{\text{ph}}(\tau)]d\tau$, where
$\gamma(t\rightarrow\infty)\propto {\rm LDOS}(\omega_x)$. McCutcheon and Nazir~\cite{Nazir}  use a similar approach to show that
$\tilde\gamma\rightarrow\gamma$ for a free-space  bath function. 

To appreciate how phonons modify the SE rate in a structured photonic reservoir,
   we first consider  a simple Lorentzian cavity  (cf.~Fig.~\ref{fig:1}(a)).
For a single cavity mode, in a dielectric with a dielectric constant $\epsilon=n_b^2$, 
\begin{align}
\label{eqA21}
J_{\text{ph}}(\omega) 
=g^2\frac{1}{\pi}\frac{\frac{\kappa}{2}}{(\omega-\omega_c)^2+(\frac{\kappa}{2})^2},
\end{align}
where $g=\left[\frac{{d}^2\omega}{2\hbar\epsilon_0n_b^2 V_{\rm eff}}\right]^{\frac{1}{2}}$
is the  QD-cavity coupling rate,
and the
QD has its dipole aligned with the cavity mode polarization
and is positioned at the field antinode. 
Defining the long-time SE rate as
$\gamma \equiv \gamma(t\rightarrow\infty)$, and similarly for $\tilde \gamma$, 
from equations~(\ref{eq:SE})-(\ref{eqA21}), we obtain
\begin{align}
\tilde{\gamma}= 2g^2 \avg{B}^2 \text{Re}\left[\int_0^{\infty} 
e^{\phi(\tau)}e^{-i\Delta_{cx}\tau-\kappa\tau/2}d\tau\right],
\label{eq:SE2}
\end{align}
where $\braket{B} = \exp[-\frac{1}{2}\phi(0)]$~\cite{Imamoglu}
and $\Delta_{cx}=\omega_c-\omega_x$.
We  can now generalize  the Purcell factor [PF] for the  enhanced
SE rate of a QD in a semiconductor cavity:
\begin{eqnarray}
{\rm PF} = \left [\frac{3}{4\pi^2}
\left ( \frac{\lambda_0}{n_b} \right )^3 \frac{Q}{V_{\rm eff}}\left(
\frac{\frac{\kappa^2}{4}}{\Delta_{cx}^2+\frac{\kappa^2}{4}}\right)\right ]
\chi,
\label{eq:PF}
\end{eqnarray}
where $\lambda_0 = \omega_c/(2\pi c)$, $Q = \omega_c/\kappa$ is the quality factor, and $\chi\equiv\tilde\gamma/\gamma$ is
the phonon-modification factor.

To obtain a mean-field approximation in the high $Q$ limit,
i.e.,  in the limit $\kappa \ll 2\pi/\tau_{\rm pn}$
(where $\tau_{\rm pn}\approx 1\,$ps is the phonon relaxation time), then $\tilde{\gamma}\rightarrow \tilde\gamma^{\text{mean}}
= \Gamma^{a^\dag \sigma^-}+2g^2 \avg{B}^2\frac{\frac{\kappa}{2}}{\Delta^2_{cx}+(\frac{\kappa}{2})^2}$, where 
the  phonon-mediated
cavity scattering rate 
$\Gamma^{a^\dagger\sigma^-}=2\avg{B}^2g^2 \text{Re}[\int_0^{\infty}d\tau e^{- i \Delta_{cx}\tau}(e^{\phi(\tau)} - 1)]$~\cite{Roy3}. This is exactly the expression for SE rate~\cite{Kaer, Roy3} which is derived (See Supplement 1, Sec. 3) with a polaron ME~\cite{Roy3} when treating the cavity mode as a system operator, with phenomenological damping $\kappa$ and using the bad cavity limit ($\kappa > g/2$).
Our theory thus not only recovers previous (polaronic) cavity-QED results
in the appropriate limit, but 
also reveals a fundamental limitation
of these approaches for sufficiently large $\kappa$ (low $Q$) cavities.
Specifically, when the cavity relaxation time becomes comparable to $\tau_{\rm pn}$, or smaller, these
formalisms break down.

\begin{figure}[t]
\vspace{0.cm}
\includegraphics[width=0.99\columnwidth, angle=0]{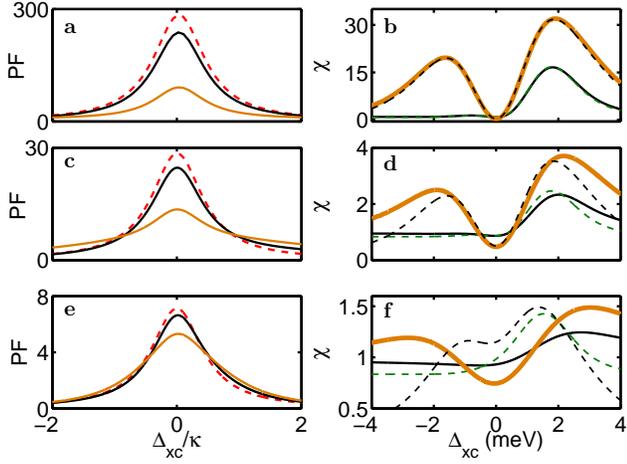}
\vspace{-0.cm}
\caption{\label{fig1} \footnotesize{ {\bf  Phonon-modified spontaneous emission of a QD in a dielectric cavity} Near-resonant PFs (left panels) and phonon-modified SE factor $\chi$ (right panels) for cavities with $\kappa =$ 0.06 meV (a, b), 0.6 meV (c, d) and 2.4 meV (e, f).
We use the continuous form of the phonon spectral function, $J_{\rm pn}(\omega) = \alpha_p\omega^3 \exp[-\frac{\omega^2}{2\omega_{ b}^2}]$, for longitudinal acoustic  phonon interaction,
and adopt experimental numbers for InAs QDs\cite{Weiler}:
 cavity-QD coupling rate  $g=0.08$ meV, phonon cutoff frequency $\omega_{ b} = 1$ meV, and exciton-phonon coupling strength  $\alpha_{ p}/(2\pi)^2 = 0.06 \rm\, ps^2$~~\cite{Roy3}.
 The dark (red) dashed line on left panels is the PF without phonons. The dark and light solid lines correspond to phonon-modified PF and $\chi$ at T = 4 K and 40 K, respectively. The dashed lines on the right panels plot $\chi^{\rm mean}$ (see text) at T = 4 K (light) and 40 K (dark).}}
\label{fig:3}
\end{figure}

\begin{figure}[t]
\vspace{0.cm}
\includegraphics[width=0.99\columnwidth]{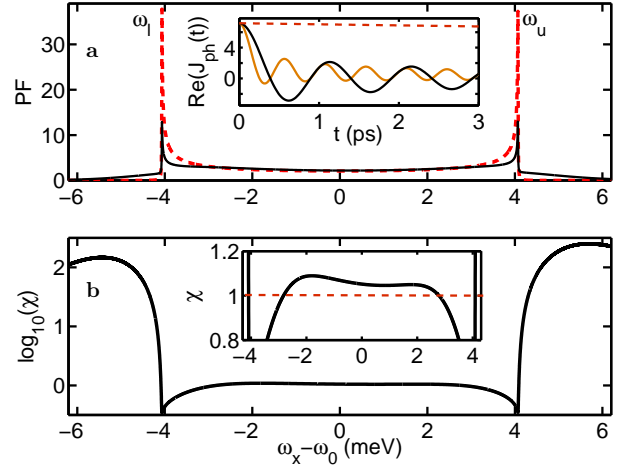} 
\vspace{-0.cm}
\caption{\label{fig4}
\footnotesize{ {\bf  Phonon-modified spontaneous emission of a QD in a coupled-cavity waveguide} (a) Purcell factors  and (b) phonon-modified SE factor $\chi$ in $\text {a log}_{10}$ scale, for a PC CROW structure, where $\omega_0$ marks the band-center. For the waveguide, PF = $\gamma/\gamma_h$, where $\gamma_h = d^2n_b\omega^3/(6\pi\hbar\epsilon_0 c^3)$ is the SE rate within the homogenous slab material. The solid (dashed) line in (a) represents PF with (without) phonons; the  inset shows $J_{\rm ph}(t)$ at the upper mode edge (light line) and band center (dark line) in arbitrary units; the dashed line shows  the simple exponential damping function $e^{-\lambda t}$ (see text).  The phonon calculations are performed at T = 40 K. Inset in (b) plots $\chi$ in a linear scale to show better display $\chi$ inside the waveguide band. }}
\end{figure}

To test this hypothesis,
 consider the example of three different cavity decay rates, 
$\kappa=0.06, 0.6,$ and $2.4$ meV, corresponding to $Q$ factor of around 23000, 2300, and 600, respectively (with $\omega_c/2\pi = 1440 $ meV).
In Fig.~\ref{fig:3} we plot the
 PF (left panels) and phonon-modified SE factor $\chi =\tilde{\gamma}/\gamma$ (right panels). For each cavity,
we investigate two different bath temperatures (4\,K and 40\,K), and the dashed lines on the left panels represent PFs without phonon modification. The clear asymmetry in $\chi$ about the LDOS peak arises since  phonon emission is more probable than absorption~\cite{Roy3} at low temperatures. The main results can be explained by writing $\tilde\gamma = \avg{B}^2\gamma+\tilde\gamma_{\rm nl}$ where $\avg{B}^2\gamma$ is the coherently renormalized bare SE rate and arises due to local ($\omega_x$) sampling of photonic LDOS, while  $\tilde\gamma_{\rm nl}= \avg{B}^2\text{Re}[2\int_0^\infty (e^{\phi(\tau)}-1)J_{\text{ph}}(\tau)]d\tau$ accounts for the {\it non-local} contribution (i.e., frequencies that would not contribute to a Fermi's golden rule expression). Note when $\kappa$ is small, $\tilde\gamma_{\rm nl} \rightarrow \Gamma^{a^\dag \sigma^-}$. Due to the non-local component, the reduction of the SE
rate is always $\geq\avg{B}^2$, at zero detuning. At large detunings, the non-local component dominates leading to an overall enhancement of the SE rate.  Figures \ref{fig:3}(b, d, f) show that $\chi$ varies significantly over several meV. The dashed lines on right panels represent $\chi^{\rm mean}
=\tilde\gamma^{{\rm mean}}/\gamma$, which evidently differs from our full calculations in the limit $\kappa\approx \tau_{\rm pn}^{-1}$ (Figs.\ref{fig:3} (d,f)). This is because  
the reservoir structure of the photon bath  is properly accounted in the present calculations  and is not approximated as a high-$Q$ cavity (also see Supplement 1, Sec 1).  It is also important to note that  low-$Q$ (several hundred)
 cavities are commonly employed for measuring the vertical emission from QDs in planar cavities~\cite{Weiler} and  for modifying  the SE rates in  simple photonic crystal cavities~\cite{Dirk}, while intermediate $Q$ ($\approx$ 3000) cavities are used for all optical switching~\cite{Edo} and enabling  single photon sources~\cite{Foell}.

We now depart from the simple cavity, and  consider the richer case of a photonic crystal CROW (cf.~Fig.~\ref{fig:1}(b)). Photonic crystal waveguides (Fig.~\ref{fig:1}(b)) can be used for realizing slow-light propagation~\cite{Notomi1} and for manipulating the emission properties of embedded QDs  for on-chip single photon emission~\cite{Rao1, Shields}, with a number of recent experiments emerging.  Current theories in this regime either ignore phonon coupling or assume a coherent renormalization factor ($\avg{B}^2\gamma$), consistent with a Fermi's golden rule (modified by a mean-field reduction factor). For the photon reservoir function, 
we  adopt a model LDOS for a  CROW~\cite{Yariv}, and use an analytical tight-binding technique to calculate the photonic band structure~\cite{Fussell1};  the photon reservoir spectral function  is obtained analytically, as
\begin{align}
J_{\text{ph}}(\omega) 
\!=\!\frac{-{d}^2\omega}{2\hbar\epsilon_0n_b^2 V_{\rm eff}} \frac{1}{\pi}
{\rm Im}\left[ \frac{1}{\sqrt{(\omega-\tilde\omega_u)(\omega-\tilde\omega_l^*)}}\right],
\end{align}
where $\tilde\omega_{u,l}=\omega_{u,l}\pm i\kappa_{u,l}$\cite{Fussell1},  $\omega_{u,l}$ is the mode-edge frequencies of the waveguide (see Fig.~\ref{fig4}(a)), $\kappa_{u,l}$ are effective damping rates, and $V_{\rm eff}$ is the mode volume of a single cavity. The photonic LDOS has a rich non-trivial spectral structure compared to a smooth Lorentzian cavity, especially within the band width of the phonon bath (which spans about 5-10 meV (See Supplement 1, Sec.~1).
 For our calculations, we use parameters that closely represent a CROW made up of a local width modulation of a line-defect photonic crystal cavity~\cite{Kuramochi} which yields a  band structure~\cite{Fussell2} consistent with experiments~\cite{Notomi2}. In Fig.~\ref{fig4}(a), we show the calculated PF  with (solid) and without (dashed) phonons, using a bath temperature of 40\,K. We see that in contrast to current theories, phonons significantly influence the spectral shape of the  SE rates, causing a reduction at the mode edges and a significant enhancement inside and outside the waveguide band. Figure \ref{fig4}(b) shows a slight asymmetry
(which increases with decreasing temperatures) in $\chi$ which is again due to unequal phonon emission and absorption rates. 
The spectral dependence of $\chi$ in a 
 the waveguide can be qualitatively understood by treating $J_{\text{ph}}(\omega)$ approximately as a sum of two Lorentzians located at the mode edges ($\omega_{u,l}$). The corresponding $J^{\rm approx}_{\text{ph}}(t) = e^{-(i\omega_u+\lambda)t}+e^{-(i\omega_l+\lambda)t}$ is a sum of two exponentially damped oscillatory functions, where $2\lambda$ is the bandwidth of the waveguide mode edge LDOS. 

At a sharp mode-edge of the LDOS, the contribution from local ($\omega_x$) photonic LDOS dominates and $\chi\approx\avg{B}^2$~\cite{Roy1}. Away from the mode-edge (Figs.~\ref{fig4}(a-b)), SE rate is enhanced due to non-local effects. This simple model discussion is, however, approximate as in reality the mode-edge LDOS is non-Lorentzian. For a symmetric Lorentzian with the same bandwidth, $\lambda^{-1}\approx 50$ ps, $J_{\text{ph}}(t)$ damps much faster than $\lambda^{-1}$ initially and damps very slowly thereafter (Fig.~\ref{fig4}(a), inset). The long time decay rate is set by the linewidth of the sharper side of the mode-edge LDOS ($\approx 0.1$ ns). This non-Lorentzian mode-edge in turn leads to a very strong enhancement of the PF ($\times 200$) outside the waveguide band (Fig.~\ref{fig4}(b)), compared to a symmetric Lorentzian line-shape (see Fig.~\ref{fig:3}(b)).



In conclusion, we have demonstrated how the frequency dependence of the LDOS of a photonic reservoir determines the extent to which phonons modify the SE of a  coupled QD. The relative dynamics between the phonon and the photon bath correlation functions is found to play a fundamentally important role; specifically, when the relaxation times are comparable, phonons strongly modify the emission spectra leading to non-Lorentzian cavity lineshapes and even enhanced SE. These effects are not obtained using the usual Fermi's golden rule.
Our formalism is important for understanding related experiments with QD-cavity systems,
such as with photoluminescence intensity measurements
with a coherent drive~\cite{OL2015}, 
and is broadly applicable to various photonic reservoir systems. 

We thank D. P. S. McCutcheon and A. Nazir for useful discussions and for sharing their results of [14]  prior to publication.

\section*{Funding Information}
Natural Sciences and Engineering Research Council of Canada.


\end{document}


\abovedisplayskip=6pt
\abovedisplayshortskip=6pt
\belowdisplayskip=6pt
\belowdisplayshortskip=6pt
\title{Supplementary material for ``Spontaneous emission from a quantum dot
in a structured photonic reservoir: phonon-mediated breakdown of Fermi's golden rule''}


\author{Kaushik Roy-Choudhury and Stephen Hughes}

\affiliation{Department of Physics, Queen's University, Kingston, Ontario, Canada, K7L 3N6}

%

\begin{abstract}
In this document we present supplementary information for our accompanying manuscript "Spontaneous emission from a quantum dot
in a structured photonic reservoir: phonon-mediated breakdown of Fermi's golden rule". First,  we show the photon and photon spectral function in the frequency and time domains, using the simple cavity structures introduced in the main text. Second, we show a derivation of the phonon-modified spontaneous emission (SE) rate, equation~(2) in  our main paper. Third, we derive the cavity-induced SE rate from a previous polaron master equations using cavity-QED equations,
and show direct agreement with our results  in the high $Q$ limit  (i.e., with $\tilde{\gamma}^{\rm mean}$ in the main paper). Last, we discuss how one can include the influence from a coherent pump field into the quantum master equation and incoherent  scattering terms, while accounting for both photon and phonon bath coupling.
\end{abstract}

\ocis{(270.0270) ; (350.4238) ; (160.6000) .}

%
%



\maketitle

\section{Cavity Photon and phonon bath functions}
\label{sec1}
\begin{figure}[h]
\vspace{0.cm}
\includegraphics[width=0.99\columnwidth]{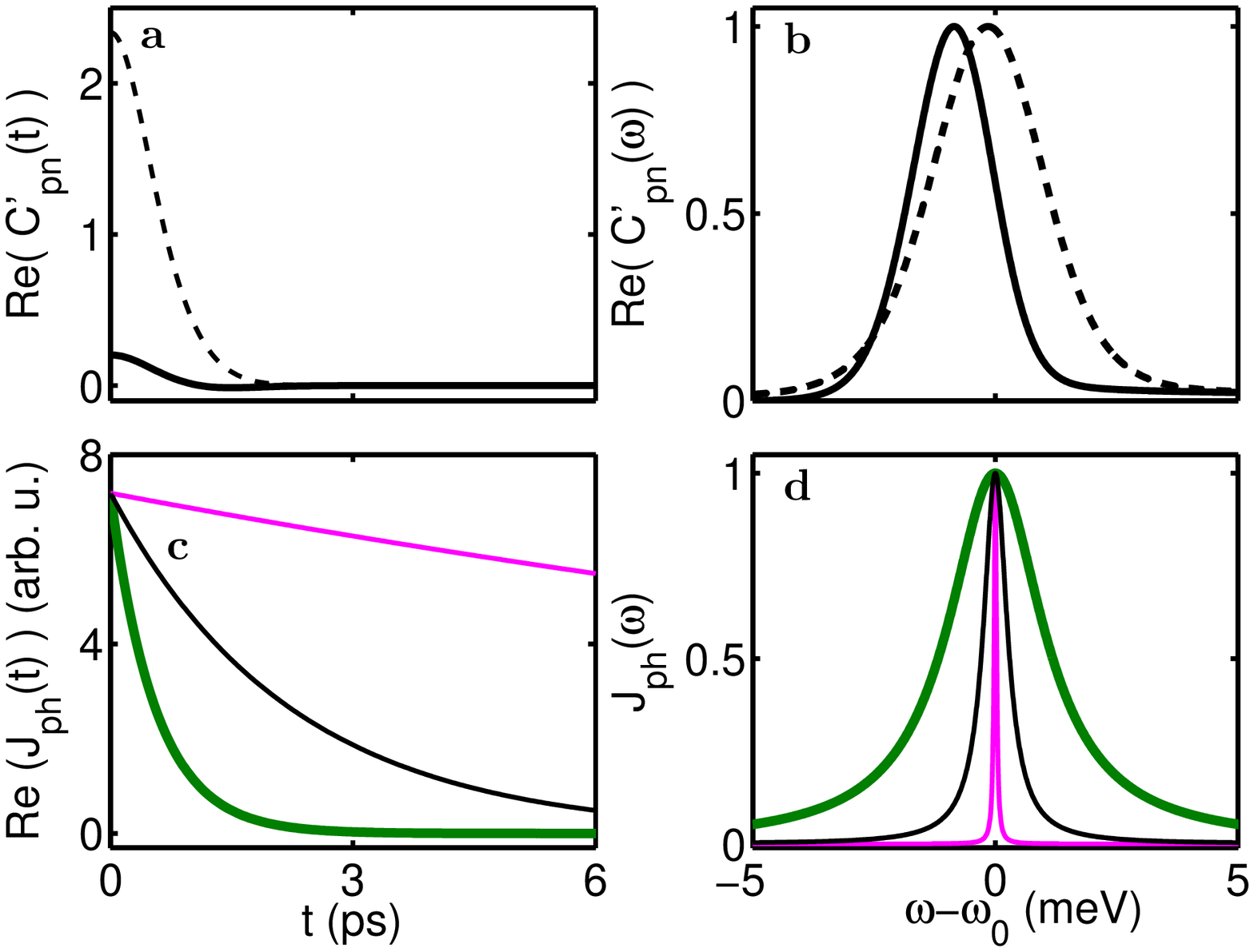}
\caption{\label{fig1} \small{ {\bf Phonon and photon bath function in time and frequency domain} ({\bf a}) Time evolution of the real part of phonon bath correlation function, $C'_{\text{pn}}(t)$ for $T = 4$ K (solid line) and $40$ K (dashed line).({\bf b}) Corresponding phonon bath function ${\rm Re}[C'(\omega)]$. ({\bf c}) Time evolution of real part of photon  bath correlation function, $J_{\text{ph}}(t)$ for cavities with $\kappa=0.06$ meV (thin light, $Q\approx 23000$), $\kappa=0.6$ meV (thin dark, $Q\approx 2300$) and $\kappa=2.4$ meV (thick dark, $Q\approx 600$) at  $\Delta_{xc} = 0$ meV. ({\bf d}) Corresponding $J_{\rm ph}(\omega)$, where $\omega_0$ is   $\omega_c$.}}
\label{fig:2}
\end{figure}

For phonon bath temperatures of 4 K and 40 K, 
Figs.~\ref{fig:2}(a) and \ref{fig:2}(b) show the phonon correlation function versus time and frequency, respectively. 
For simplicity, a polaron shift $\Delta_P$ (defined below in Section \ref{sec2}) is implicitly absorbed in the definition of $\omega_{x}$, and we  also define a phonon correlation function that decays to zero, through
$C'_{\text{pn}}(\tau)=e^{\phi(\tau)}-1$.
All phonon parameters  and $\phi(\tau)$ are given in our main paper.
The time evolution of the real part of the phonon correlation function  shows that typical phonon correlation times are very fast ($\tau_{\rm pn}\leq 3$ ps). For comparison,
we also show a photon correlation function in
Figs.~\ref{fig:2}(c) and \ref{fig:2}(d), for the three
different values of $\kappa$ used in our  paper:
$\kappa = 2.4$ meV (thick dark line), 0.6 meV (thin dark line), and 0.06 meV (thin light line).
 The cavity bath correlation functions $J_{\text{ph}}(t)$ are oscillatory functions damped at the cavity decay rate, with an  oscillation frequency is determined by the QD-cavity detuning 
$\Delta_{xc}$, and we show the case for zero detuning.  From equation~(2) in our  paper, and Fig.~\ref{fig:2} below, we expect  that  ($i$) phonons should not influence the SE rate $\gamma$ in a strongly damped cavity and ($ii$) phonons will reduce the SE rate to its mean-field limit~\cite{Roy} ($\avg{B}^2\!\gamma$) only for a weakly damped cavity.
   Significant deviations from these two limits
occur  when the damping times of the photon and phonon correlation functions are comparable. This is shown explicitly in Fig.~2 of the main text using the cavity spectral functions shown below.
It is also shown explicitly in Fig.~3 for the slow-light
photonic crystal waveguide.

\section{Derivation of phonon-modified spontaneous emission rate}
Theoretical descriptions of electron-phonon scattering in QDs systems range from the independent Boson model~\cite{Axt2}, correlation expansion~\cite{knorr}, perturbative master equations (MEs) \cite{Nazir1}, to polaron MEs~\cite{Roy2, Imamoglu,Hohenester,Kaer}, variational~MEs \cite{Nazir3} and real-time path integrals~\cite{Axt}. In our Letter we
introduce a self-consistent ME approach with both phonon and photon reservoirs included.
We model the QD as a two-level system interacting with an inhomogeneous semiconductor based photonic reservoir and an acoustic phonon bath. Assuming a QD of dipole moment $\v{d} = d\v{n_d}$ and position at $\r1_d$, the total Hamiltonian of the coupled system in a frame rotating at the QD exciton frequency $\omega_x$, can be written as~\cite{Scheel}
\begin{align}
\label{eqA1}
    H &= \hbar\int d\v{r} \int_0^{\infty}d\omega\,\fdag(\r1,\omega)\f(\r1,\omega)  
-[\smdag e^{i\omega_xt}\int_0^{\infty}d\omega\,\v{d}\cdot\v{E}(\r1_d,\omega) + \text{H.c}]  
+ \Sigma_q\hbar \omega_q b^{\dag}_qb_q + \smdag\sm\Sigma_q\hbar\lambda_q(b^{\dag}_q+b_q),
\end{align}
where $\smdag$/$\sm$ are the Pauli operators of the exciton 
(electron-hole pair), $b_q$/$b^{\dag}_q$ are the annihilation and creation operators of the acoustic phonon reservoir, and the exciton-phonon coupling strength $\lambda_q$ is assumed to be real. The operators $\f$/$\fdag$ are the boson field operators of the photon reservoir and these satisfy the usual commutation rules for boson operators. The interaction between the QD and the photonic reservoir is written using the dipole and the rotating wave approximation. The electric-field operator $\v{E}(\r1,\omega)$ is given by $\v{E}(\r1,\omega) =i\int d\r1' \v{G}(\r1,\r1';\omega)\sqrt{\frac{\hbar}{\pi\epsilon_0}\epsilon_I(\r1',\omega)}\f(\r1',\omega)$~\cite{Scheel}, where $\v{G}(\r1,\r1';\omega)$ is the electric field Green's function for the medium and $\v{E}(\r1,\omega)$  satisfies the Kramers-Kronig relations, with a complex dielectric constant $\epsilon=\epsilon_R+i\epsilon_I$.

To include phonon interactions nonperturbatively, we  perform the polaron transform on the Hamiltonian $H$ given by $H' \rightarrow e^P He^{-P}$, where $P = \smdag\sm \Sigma_q\frac{\lambda_q}{\omega_q} (b^{\dag}_q-b_q)$~\cite{Mahan,Imamoglu}. $H'$ consists of the following reservoir and interaction terms, 
\begin{align}
\label{eqA2}
H' =& H'_{\text{R}}+H'_{\text{I}}\nn,\\
H'_{\text{R}} =& \hbar\int d\v{r} \int_0^{\infty}d\omega\,\fdag(\r1,\omega)\f(\r1,\omega)+\Sigma_q\hbar \omega_qb^{\dag}_qb_q\nn,\\
H'_{\text{I}} =&   -[B_+\smdag e^{i\omega_xt}\int_0^{\infty}d\omega\,\v{d}\cdot\v{E}(\r1_d,\omega) + \text{H.c}], 
\end{align}
\noindent where the coherent phonon bath displacement operators $B_{\pm}$ are defined as $B_{\pm} = \exp[\pm\Sigma_q\frac{\lambda_q}{\omega_q}(b_q-b^{\dag}_q)]$. A polaron frequency shift, $\Delta_P = \int_0^{\infty}d\omega\frac{J_{\rm pn}(\omega)}{\omega}$  is implicitly absorbed in the definition of $\omega_x$, where $J_{\rm pn}(\omega)$ is the phonon spectral function~\cite{Roy3}. For our calculations, we use the continuous form of the phonon spectral function $J_{\rm pn}(\omega) = \alpha_p\omega^3 \exp[-\frac{\omega^2}{2\omega_{ b}^2}]$ for longitudinal acoustic (LA) phonon interaction, resulting primarily from deformation potential coupling in InAs QDs~\cite{Ulhaq}. In the above formula, $\omega_{ b}$ is the phonon cutoff frequency and $\alpha_{ p}$ is the exciton-phonon coupling strength~\cite{Roy3}. 

We subsequently transform the Hamiltonian $H'$ to the interaction picture using $\tilde{H'} \rightarrow U^{\dag}(t)H'U(t)$ with $U(t) = \exp[-iH'_{\text{R}}t/\hbar ]$. Assuming weak coupling with the photon bath, a time-convolutionless~\cite{Breuer} polaron master equation for the QD reduced density operator $\rho$ is then derived using second-order Born approximation in $\tilde{H'_{\text{I}}}$. The following interaction-picture ME is ontained,
\begin{align}
\label{eqA3}
\frac{\partial \tilde{\rho}(t)}{\partial t} = 
-\frac{1}{\hbar^2}\int_0^t d\tau \text{Tr}_{\text{R}_{\text{ph}}} \text{Tr}_{\text{R}_{\text{pn}}}\{ [\tilde{H'}_{\text{I}}(t),[\tilde{H'}_{\text{I}}(t-\tau),\tilde{\rho}(t)\rho_{\text{R}}]]\},
\end{align}
where $\tilde{\rho}$ is the reduced density operator of the QD in the interaction picture and  $\text{Tr}_{\text{R}_{\text{ph(pn)}}}$ denotes trace with respect to the photon (phonon) reservoirs which are assumed to be statistically independent, 
$\rho_{\text{R}} = \rho_{\text{R}_{\text{ph}}}\rho_{\text{R}_{\text{pn}}}$~\cite{Carmichael}. The density operator of the photonic reservoir $\rho_{\text{R}_{\text{ph}}}$ is assumed to be initially in thermal equilibrium. We use the bath approximation, $\text{Tr}_{\text{R}_{\text{ph}}}[\f(\v{r},\omega),\fdag(\v{r'},\omega')] = [\tilde{n}(\omega)+1]\delta(\v{r}-\v{r'})\delta(\omega-\omega')$ and 
$\text{Tr}_{\text{R}_{\text{ph}}}[\fdag(\v{r},\omega),\f(\v{r'},\omega')] = \tilde{n}(\omega)\delta(\v{r}-\v{r'})\delta(\omega-\omega')$ and consider the zero temperature limit ($\tilde{n}(\omega) = 0$). Using the relation $\int d\v{s}\epsilon_I(\v{s},\omega)\v{G}(\r1,\v{s},\omega)\v{G}^*(\v{s},\r1';\omega) = \text{Im}[\v{G}(\r1,\r1';\omega)]$, transforming to the Schr\"{o}dinger picture, and carrying out the trace over the photon~\cite{Tanas}  and phonon reservoir, we derive the following generalised ME: $\frac{\partial \rho}{\partial t} = \mathcal{L}_{\text{ph}}(\rho)$, where
\begin{align}
\label{eqA5}
\mathcal{L}_{\text{ph}}(\rho)&= \int_0^t d\tau\! \int_0^{\infty}\! d\omega\,  {J_{\text{ph}}(\omega)}[-C_{\text{pn}}(\tau)\smdag\sm e^{i\Delta_x\tau}\rho 
+C^*_{\text{pn}}(\tau)\sm\rho\smdag e^{-i\Delta_x\tau}
+C_{\text{pn}}(\tau)\sm\rho\smdag e^{i\Delta_x\tau} \nn\\&- C^*_{\text{pn}}(\tau)\rho\smdag\sm e^{-i\Delta_{x}\tau}],
\end{align}
and $\Delta_x = \omega_x-\omega$. The photon-reservoir spectral function $J_{\text{ph}}(\omega)$, is given by $J_{\text{ph}}(\omega)= \frac{\v{d}\cdot \text{Im}[\v{G}(\r1_d,\r1_d;\omega)]\cdot \v{d}}{\pi\hbar\epsilon_0}$ and the phonon bath correlation function $C_{\text{pn}}(\tau)$ is defined as $C_{\text{pn}}(\tau)=e^{[\phi(\tau)-\phi(0)]}$, where
$\phi(t) = \int_0^{\infty} d\omega\frac{J_{\rm pn}(\omega)}{\omega^2}[\coth(\hbar\omega/2k_BT)\cos(\omega t)- i\sin(\omega t)]$. A simple expression for the phonon-modified SE decay rate can be derived from the real part of $\mathcal{L}_{\text{ph}}$, so that $\text{Re}(\mathcal{L}_{\text{ph}})$ reduces to the Lindblad form, $\tilde{\gamma}(t) L[\sigma^-]$, where $\tilde{\gamma}$ is the phonon-modified SE decay rate of a QD given by 
\begin{align}
\label{eqA6}
\tilde{\gamma}(t) = 2\int_0^t\text{Re}[C_{\text{pn}}(\tau)J_{\text{ph}}(\tau)]d\tau,
\end{align}
where $J_{\text{ph}}(\tau) = \int_0^{\infty} d\omega J_{\text{ph}}(\omega) e^{i(\omega_x-\omega)\tau}$ is the photon bath correlation function, and ${L}[O]
=\frac{1}{2}(2O\rho O^\dagger -O^\dagger O \rho-\rho O^\dagger O)$. Note that the imaginary part of $\mathcal{L}_{\text{ph}}$ yield Lamb shifts~\cite{Roy3}. This is Equation~(2) in our main paper.  In absence of the phonon coupling, the SE decay rate of the QD in a structured photon reservoir reduces to $\gamma(t) = 2\int_0^t\text{Re}[J_{\text{ph}}(\tau)]d\tau$. A similar expression for $\tilde{\gamma}$ was derived by McCutcheon and Nazir~\cite{Nazir}, who then use a free space photon reservoir function  to show no phonon-modification to the SE rate.\\

\label{sec2}

\section {Correspondence with previous work in the mean field cavity-QED regime: Derivation of $\tilde{\gamma}^{\rm mean}$ using a polaron cavity-QED master equation}

Using the  polaron transformed effective Lindblad master equation of an coupled QD-cavity system~\cite{Roy3}, 
an expression for the SE rate can be derived in weak excitation approximation (WEA) and bad cavity limit. The effective phonon master equation is defined through,
\begin{align}
\label{eqA7}
\frac{\partial \rho}{\partial t} = \frac{1}{i\hbar}[H^{\rm eff}_{\rm sys},\rho] + \Gamma^{a^\dag \sigma}L(a^\dag \sigma^-) + \Gamma^{\sigma^+ a }L(\sigma^+ a)+\kappa L(a),
\end{align}
where $a$ is the cavity lowering operator and $H^{\rm eff}_{\rm sys} = \hbar\Delta_{cx} a^{\dagger}a+\hbar\avg{B} g(a^\dag \sigma^-+\sigma^+a)$, in a frame rotating at the exciton frequency $\omega_x$. The QD-cavity coupling strength is $g$ and the cavity damping rate is $\kappa$ and $\avg{B} = \avg{B_+} = \avg{B_-}$ is
the thermally averaged phonon bath dispacement operator~\cite{Mahan}.
 The cavity (exciton) feeding terms are $\Gamma^{a^\dag \sigma^-/\sigma^+ a}=2\avg{B}^2g^2 \text{Re}[\int_0^{\infty}d\tau e^{\mp i \Delta_{cx}\tau}(e^{\phi(\tau)} - 1)]$~\cite{Roy3}.

In the weak excitation approximation, the Bloch equations for $\avg{a}$ and $\avg{\sigma^-}$ derived from Equation~(\ref{eqA7}) are given by
\begin{align}
\label{eqA9}
\frac{d\avg{a}}{dt}=& = - \left (i\Delta_{cx}+\frac{\kappa+\Gamma^{\sigma^+ a}}{2}\right)\avg{a}-i\avg{B}g\avg{\sigma^-}, \nn\\
\frac{d\avg{\sigma^-}}{dt}=& =  - \frac{\Gamma^{a^{\dagger} \sigma}}{2}\avg{\sigma^-}-i\avg{B}g\avg{a}.
\end{align}
The coupled equations can be expressed in a matrix form as $\v{V}=\v{MV}$, where $\v{V} = [\avg{a}; \avg{\sigma}]$. In the bad cavity limit, the real part of the eigenvalues of $\v{M}$ give the decay rates $\avg{a}$ and $\avg{\sigma}$. The SE rate of the QD is twice the decay rate of $\avg{\sigma^-}$ and is given by
\begin{align}
\label{eqA10}
&\tilde{\gamma}_{\rm WEA} = {\text Re} \Bigg[ -\left(\frac{\Gamma^{a^{\dagger} \sigma^-}+\kappa_1}{2}+i\Delta_{cx}\right) -\sqrt{\left(\frac{\Gamma^{a^{\dagger} \sigma^-}+\kappa_1}{2}+i\Delta_{cx}\right)^2-4\left (\avg{B}^2g^2+\frac{\Gamma^{a^{\dagger} \sigma^-}}{2}(\frac{\kappa_1}{2}+i\Delta_{cx})\right )}\Bigg],
\end{align}
where $\kappa_1 = \kappa+\Gamma^{\sigma^+ a}$. The above expression can be manipulated to give
\begin{align}
\label{eqA11}
&\tilde{\gamma}_{\rm WEA} = {\text Re} \Bigg[-\left(\frac{\Gamma^{a^{\dagger} \sigma^-}+\kappa_1}{2}+i\Delta_{cx}\right)-\sqrt{\left(\frac{\Gamma^{a^{\dagger} \sigma^-}-\kappa_1}
{2}-i\Delta_{cx}\right)^2-4\avg{B}^2g^2}\Bigg],
\end{align}
which, in the bad cavity limit ($\kappa \gg g/2$), can be further simplified by expanding the second term to the first power of $\frac{4\avg{B}^2g^2}
{\left(\frac{\Gamma^{a^{\dagger} \sigma^-}-\kappa_1}{2}-i\Delta_{cx}\right)^2}$. The final simplified expression for the QD SE rate is
\begin{align}
\label{eqA11}
\tilde{\gamma}_{\rm WEA}^{\rm bad} = \Gamma^{a^{\dagger} \sigma^-}+2\avg{B}^2g^2
\frac{\left(\frac{\kappa +\Gamma^{\sigma^+ a}
-\Gamma^{a^{\dagger} \sigma^{-}}}{2}\right)}{\Delta_{cx}^2+
\left(\frac{\kappa +\Gamma^{\sigma^+ a}-\Gamma^{a^{\dagger} \sigma^-}}{2}\right)^2}.
\end{align}
 Moreover, for $\kappa \gg\Gamma^{\sigma^+ a}-\Gamma^{a^{\dagger} \sigma^-}$,
\begin{align}
\label{eqA11}
\tilde{\gamma}_{\rm WEA}^{\rm bad} = \Gamma^{a^{\dagger} \sigma^-}+2\avg{B}^2g^2\frac{(\frac{\kappa}{2})}{\Delta_{cx}^2+(\frac{\kappa}{2})^2},
\end{align}
which is equivalent to $\tilde{\gamma}^{\rm mean}$ in our paper.
Using the same polaronic cavity-QED approach,
an identical expression is derived in Ref.~\cite{Kaer} in the large detuning limit ($\Delta  \gg g$) by adiabatically eliminating the cavity. Our paper makes it clear that $\tilde{\gamma}_{\rm WEA}^{\rm bad}$, and indeed the entire polaron cavity-QED master equation, is only valid when $\kappa \ll 2\pi/\tau_{\rm pn}$ ($\tau_{\rm pn}$ is around 1\, ps for InAs QDs).

\section{Comments about how to INCLUDE a coherent pump field}
 The QD can be excited using a weak coherent pulse of the form $H_{\rm pump} =\eta_x(\sigma^+ e^{-i\omega_Lt}+\sigma^-e^{i\omega_Lt})$, where $\eta_x$ and $\omega_L$ are the amplitude and central frequency of the pump pulse. For resonant excitation $\omega_L = \omega_x$. Under this situation, the polaron-transformed system Hamiltonian, $H'$ (Equation~(\ref{eqA2})) has the form
\begin{align}
\label{eqA12}
H' =& H'_{\text{S}}+H'_{\text{R}}+H'_{\text{I}}\nn,\\
H'_{\text{S}}=& \hbar \avg{B}\eta_x [\sigma^++\sigma^-]\nn,\\
H'_{\text{R}} =& \hbar\int d\v{r} \int_0^{\infty}d\omega\,\fdag(\r1,\omega)\f(\r1,\omega)+\Sigma_q\hbar \omega_qb^{\dag}_qb_q\nn,\\
H'_{\text{I}} =&   -[B_+\smdag e^{i\omega_xt}\int_0^{\infty}d\omega\,\v{d}\cdot\v{E}(\r1_d,\omega) + \text{H.c}]+X_g\zeta_g +X_u\zeta_u, 
\end{align}
where $\zeta_g = \frac{1}{2}(B_+ +B_-- 2\avg{B})$ and $\zeta_u =\frac{1}{2i}(B_+-B_-)$ are the phonon induced fluctuation operators~\cite{Imamoglu} and $X_g$ and $X_u$ are defined through $X_g = \hbar\eta_x(\sm+\smdag)$ and $X_u = i\hbar\eta_x(\smdag-\sm)$. Following the same steps as before (Section II), we subsequently transform the Hamiltonian $H'$ to the interaction picture using $\tilde{H'} \rightarrow U^{\dag}(t)H'U(t)$ where $U(t) = \exp[-i(H'_{\text{S}}+H'_{\text{R}})t/\hbar ]$. Assuming weak coupling with the photon bath, a time-convolutionless~\cite{Breuer} polaron master equation for the QD reduced density operator $\rho$ is then derived using second-order Born approximation in $\tilde{H'_{\text{I}}}$. After carrying out the trace over the photon and phonon baths, the final ME for the QD reduced density operator $\rho$ in the Schr\"{o}dinger picture is given by $\frac{d \rho}{ dt} =  \frac{1}{i\hbar}[H'_{\text{S}},\rho]+\mathcal{L}_{\text{ph}}(\rho) + \mathcal{L}_{\text{pn}}(\rho)$. The phonon part $\mathcal{L}_{\text{pn}}(\rho)$ remains decoupled from the photon reservoir and, to a very good approximation, can be expressed through the following Lindblad terms (enhanced radiative decay and incoherent excitation~\cite{Roy3}):
\begin{align}
\label{eqA13}
\mathcal{L}_{\text{pn}}(\rho) = \Gamma^{\sm}L[\sm] + \Gamma^{\smdag}L[\smdag]
\end{align}
where, ${\Gamma^{\sigma^{+/-}}}=2\avg{B}^2\eta_x^2 \text{Re}[\int_0^{\infty}d\tau e^{\pm i (\omega_L-\omega_x)\tau}(e^{\phi(\tau)} - 1)]$. The photon part $\mathcal{L}_{\text{ph}}$ on the other hand is modified by the phonons and is given by
\begin{align}
\label{eqA14}
\mathcal{L}_{\text{ph}}(\rho) &= \int_0^t d\tau \int_0^{\infty} d\omega\, J_{\text{ph}}(\omega) [-C_{\text{pn}}(\tau)\smdag\sm(-\tau)e^{i\Delta_x\tau}\rho + C^*_{\text{pn}}(\tau)\sm\rho\smdag(-\tau)e^{-i\Delta_x\tau}
\nn\\&+C_{\text{pn}}(\tau)\sm(-\tau)\rho\smdag e^{i\Delta_x\tau} 
- C^*_{\text{pn}}(\tau)\rho\smdag(-\tau)\sm e^{-i\Delta_x\tau}], 
\end{align}
where the time-dependent operators $\sigma^{\pm}(-\tau) = e^{-iH'_{\text{S}}\tau/\hbar} \sigma^{\pm}e^{iH'_{\text{S}}\tau/\hbar}$. This indicates that the scattering rates are pump-field dependent in general and for strong pumps, different dressed states ($\omega = \omega_x,\omega_x\pm2\eta_x$) can sample different regions of the photonic LDOS~\cite{Roy1,Roy2,Nazir3}. Such behavior leads to asymmetric Mollow triplets~\cite{John2} in the absence of phonons, if the Rabi strength is large enough to sample difference values and an asymmetric LDOS (see Ref.~\cite{Ge2013} for detailed discussions). Since we assume weak excitation, only the photonic LDOS around $\omega_x$ is sampled. Thus $\sigma^{\pm}(-\tau)$ is replaced by $\sigma^{\pm}$, which reduces $\mathcal{L}_{\text{ph}}(\rho)$ to Equation~\eqref{eqA5}, the form used in the main manuscript. The QD can also be excited by a weak incoherent pump by using higher states of the QD. The precise form of the pump is not important for the study of SE\cite{Kaer}, as long as it weakly excites the system. However, one can add in a pump field using our general approach as we show above.